\documentclass[pre, reprint,  eprintnumbers]{revtex4-2}
\usepackage[]{graphicx}
\usepackage{epstopdf}
\usepackage{bm}
\usepackage{hyperref}

\begin{document}
\title{Critical Slowing Down at the Abrupt Mott Transition: When the First-Order Phase Transition Becomes Zeroth-Order and Looks Like Second-Order}
\author{Satyaki Kundu}
\affiliation{Indian Institute of Science Education and Research Kolkata, Mohanpur, Nadia 741246, West Bengal, India}
\author{Tapas Bar}
\affiliation{Indian Institute of Science Education and Research Kolkata, Mohanpur, Nadia 741246, West Bengal, India}
\author{Rajesh Kumble Nayak}
\affiliation{Indian Institute of Science Education and Research Kolkata, Mohanpur, Nadia 741246, West Bengal, India}
\author{Bhavtosh Bansal}\email{bhavtosh@iiserkol.ac.in}
\affiliation{Indian Institute of Science Education and Research Kolkata, Mohanpur, Nadia 741246, West Bengal, India}
\eprint{Physical Review Letters {\bf 124}, 095703 (2022)}
\begin{abstract}
We report that the thermally-induced Mott transition in vanadium sesquioxide shows critical-slowing-down and enhanced variance ('critical opalescence') of the order parameter fluctuations measured through low-frequency resistance-noise spectroscopy. Coupled with the observed increase of also the phase-ordering time, these features suggest that the strong abrupt transition is controlled by a critical-like singularity in the hysteretic metastable phase. The singularity is identified with the spinodal point and is a likely consequence of the strain-induced long-range-interaction.
\end{abstract}
\maketitle
Despite their ubiquity across systems and scales \cite{qcd-LIGO, binder_rpp,McWhan_V2O3, AKR_VO2, liu, wang_schuller, del Valle, Peil, Sambandamurthy_NdNiO3, AKR_NdNiO3, Post, Manganite_singh, Chandni, Chaddah, Ong_CDW, zhu}, first-order or abrupt phase transitions typically get only a passing reference in a traditional statistical physics course \cite{Huang}. This is because they lack many of the remarkable features of the second-order or continuous transitions which arise from the diverging susceptibility \cite{Huang, chaikin-lubensky, Goldenfeld}. The singularity leads to a power-law-divergence of the correlation length, thereby leading to a description that is universal and largely independent of the microscopic details \cite{Goldenfeld, Huang, chaikin-lubensky}. Experimentally, this singularity manifests in the sharp enhancement of the dynamical time scales (critical slowing down) \cite{hartman_SlowingDownMott, chaikin-lubensky} and  the variance of fluctuations (critical opalescence) \cite{CriticalOpalescence_Mott, Chen-Yu, chaikin-lubensky}.

\begin{figure}[!b]
\begin{center}
\includegraphics[scale=0.38]{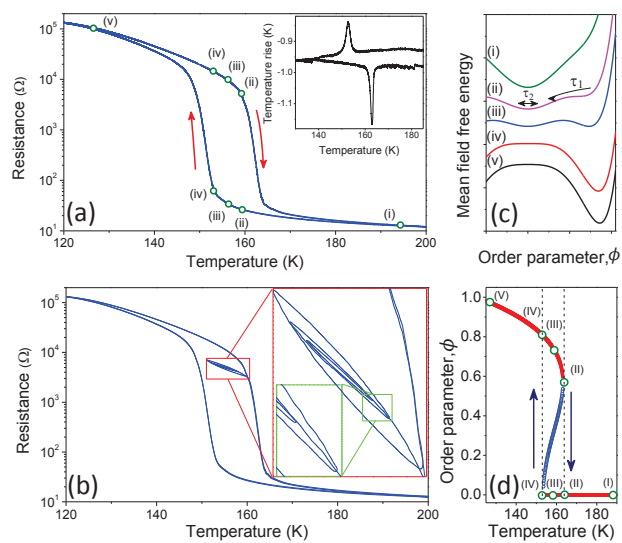}
\caption{(a) Temperature dependence of resistance of V$_2$O$_3$ shows a sharp and hysteretic transition around $153$ K while cooling and $162$ K while heating. (inset) A large latent heat is also measured in differential thermal analysis \cite{Tapas}. (b) Return-point memory \cite{Ong_CDW, Keim_RMP, sethna}: Although multivalued in the hysteretic region, after any excursion out of a given hysteresis loop in the form of a minor loop the same value of the resistance is recovered on return. Thus the memory of the excursion is wiped out. (c) Mean field free energy as a function of the spatially averaged order parameter $\phi$ for the compressible Ising model (essentially the $\phi^6$ theory) \cite{Tapas} can be used to capture transition semi-quantitatively. The points corresponding to curves (i)-(v) are also roughly marked on the resistance curve in (a). Curve (iii) represents the binodal where the two minima are equal and (ii) and (iv) are the two spinodals where one minimum becomes an inflection point. $\tau_2$ is the relaxation time of fluctuations measured through the autocorrelation of the resistance noise and $\tau_1$ is the time associated with phase-ordering. (d) Location of the extrema of the above free energy as a function of temperature with the same points (i)-(v) marked. In the temperature window $\sim 153-162$ K where hysteresis is seen, the order parameter $\phi$ has a globally stable, a metastable and an unstable extremum.}
\end{center}
\end{figure}

In this letter, we experimentally demonstrate slowing down and enhancement of fluctuations at an abrupt phase transition (APT) and argue that a certain class of APTs are also controlled by critical-like singularities. Specifically, we have studied the celebrated Mott transition in V$_2$O$_3$ \cite{imada_rmp, McWhan_V2O3, Tapas, mcleod, Kalcheim}. As the scope of this work transcends the microscopic and material details of the system under investigation, here we only highlight the two essential characteristics \cite{fn:Mott}. Firstly, that this is an APT is unambiguously inferred from the large latent heat [Fig. 1 (a) (inset)]. Second is a curious feature that---V$_2$O$_3$ shares with some other vanadates \cite{AKR_VO2, liu, wang_schuller, del Valle}, nickelates \cite{Peil, Sambandamurthy_NdNiO3, AKR_NdNiO3, Post}, manganites \cite{Manganite_singh}, intermetallic alloys \cite{Chandni, Chaddah}, other charge-ordered materials \cite{Ong_CDW, zhu}, and spin-transition polymers \cite{Kahn_Martinez, Miyashita_PRB2009,Miyashita-Konishi-Nishino-Tokoro-Rikvold}---the transition is {\em always} hysteretic.

Hysteresis indicates that this `mixed-order' behavior (viz. the observation of critical slowing down as well as a latent heat at the transition)  \cite{FN: Mixed-order, alert} may originate from the spinodal singularity \cite{binder_rpp,Stephanov, Procaccia, Sornette, Nandi, klein-monette, Saito, Gunton-Yalabik,zhong_prl2005, Sasaki, ikeda, Compagner, Chomaz, Penrose-Lebowitz}. This would then be a  manifestation of the classic mean-field physics, already contained in van der Waal's equation (without Maxwell correction) \cite{binder_rpp, klein-monette, Penrose-Lebowitz}. Within equilibrium thermodynamics, an APT occurs at the binodal, the point where the free energy minima for the two phases have the same value [Fig. 1 (c)]. But nucleation barriers can lead to supersaturation into a metastable phase \cite{Debenedetti}. The depth of supersaturation is thus a function of the efficacy of the fluctuations in affecting first-passage to the lower energy equilibrium phase \cite{Gilmore}. It is believed that only in the (zero-temperature or infinite-range-interaction) mean-field limit can the hysteretic passage though the metastable phase extend up to the stability limits, the spinodals \cite{binder_rpp, Penrose-Lebowitz}. As $\chi_T^{-1}={\delta^2 F\over \delta \phi^2}=0$ at the spinodals (where $F[\varphi]$ is the analytically continued free energy density, $\chi_{_T}$ is the susceptibility \cite{Goldenfeld}, and $\phi$ is the spatially averaged order parameter), the correlation length and relaxation times diverge at the spinodals for the same reason that they do at the critical point \cite{binder_rpp, Stephanov, Procaccia}. Consequently, the spinodals are fixed points under the renormalization group and should display critical behavior and universality \cite{Saito, Gunton-Yalabik,zhong_prl2005}.

While this instability has been studied in a variety of zero-temperature noise-free systems \cite{Procaccia,Sornette, Spinodal_rydbergGases, Jung_Gray_Roy_Mandel, Nandi, Strogatz}, spinodals have not been experimentally established in thermodynamic systems where fluctuations are present \cite{FN:Spinodal_lightScattering}. From the scaling behavior of the dynamic hysteresis and the average qualitative nature of the phase ordering, we have recently proposed that the Mott transition in V$_2$O$_3$ does indeed occur around such bifurcation points \cite{Tapas}. The present work, by focussing on fluctuations, transcends the mean-field picture to establish that the transition has a genuine thermodynamic character. Our observations of critical slowing down \cite{hartman_SlowingDownMott} and enhancement \cite{Sasaki} of the order parameter fluctuations indicate that one can at least get close enough to the spinodal such that the singularity controls many of the features of the transition, even if there are no divergences \cite{Stephanov, Gulbahce}.

On account of the criticality and the essential role of hysteresis, it is meaningful to distinguish these transitions from the usual first-order transitions \cite{binder_rpp}. Following the Ehrenfest's criterion,  these may be called `zeroth-order', as the free energy is itself discontinuous \cite{Gilmore, Tapas} [Fig 1 (c)].

\begin{figure}[!t]
\begin{center}
\includegraphics[scale=0.5]{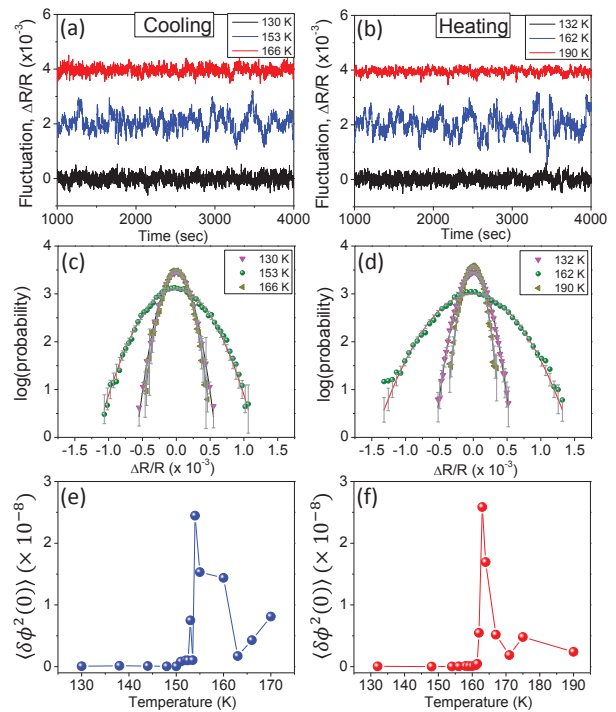}
\caption{{\em Time series of the resistance fluctuations} at different temperatures. The left column depicts the measurements done under cooling and the right panel under heating. (a, b) Normalized resistance fluctuations after the time series were detrended by imposing a low frequency cut-off of about $10^{-3}$ Hz \cite{SM}. (c, d) The probability distribution of fluctuations. Solid lines represent the Gaussian distribution with zero mean and the given variance and the error bars roughly represent $99.7\%$ confidence for this distribution. (e, f) Variance of the order parameter fluctuations $\langle \delta\phi^2(0)\rangle$ roughly inferred from the resistivity time series. The fluctuations are expected to follow the divergence of $\chi_T$ around the spinodals [Eq. 3].}
\end{center}
\end{figure}

\begin{figure}[!t]
\begin{center}
\includegraphics[scale=0.33]{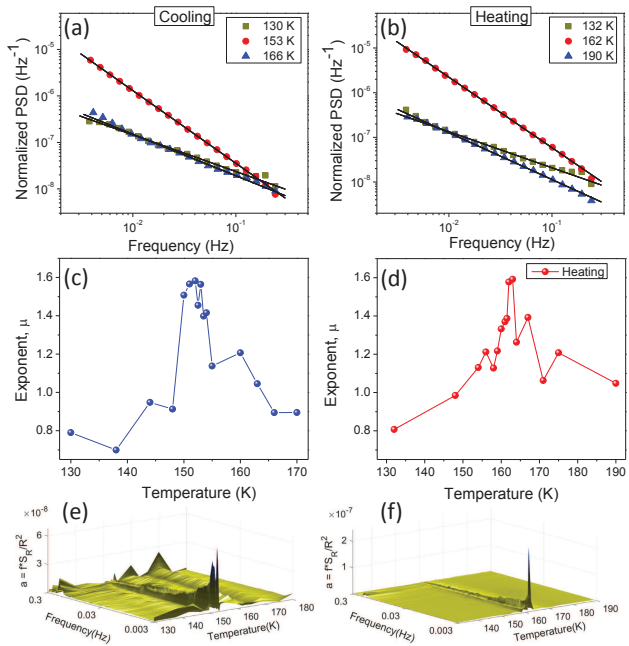}
\caption{{\em Critical-slowing-down} of fluctuations around the spinodals. (a, b) Normalized noise power spectral density (PSD), $S_R/R^2$ at some representative  temperatures. (c, d) Temperature dependence of the exponent $\mu$, where $S_f\sim f^{-\mu}$. (e, f) A surface plot of $f\times S_R/R^2$ to visualize the non-$1/f$ behavior when the spectral weight shifts to lower frequencies.}
\end{center}
\end{figure}

V$_2$O$_3$ is our material of choice \cite{Tapas} because of the sharp and definitive APT with a large latent heat over a narrow temperature window [Fig. 1 (a)]. Although one cannot define state variables in the metastable phase, the system does show reproducible quasistationary behavior (essential for the fluctuation spectroscopy), as seen in the return-point memory \cite{Ong_CDW, Keim_RMP} of hysteresis [Fig 1 (b)].

The temperature dependence of the resistance of a typical polycrystalline sample V$_2$O$_3$ used in the study, after it has been thermally cycled between $\sim 77-250$ K a few hundred times, is shown in Fig 1 (a), with the inset showing the latent heat peaks in the differential thermal analysis peaks measurement \cite{Tapas}. The hysteresis window ($\sim 153-162$ K) can be seen to contain the metastable phase whose properties are history-dependent, whereas  the regions outside this widow seem to represent equilibrium states. In rest of this Letter, results for the heating and the cooling transitions are shown in parallel \cite{SM} as the similarity of the two data sets is central to main conclusions of the work.

Resistance-noise spectroscopy has been a powerful tool for the study of fluctuations close to phase transitions \cite{hartman_SlowingDownMott, AKR_NdNiO3, Sambandamurthy_NdNiO3, Chandni}. The experiments reported here involved carefully recording the time series of the fluctuating resistance $R$ with the sample temperature kept precisely fixed at the given value \cite{SM}.

Figure 2 (a, b) shows some of these time series of the normalized resistance fluctuations $\Delta R(t)/R$ at different temperatures after the removal of the very slowly varying smooth background \cite{SM}. It is hard to discern any significant departure from their Gaussianity [Fig. 2 (c, d)]. To make contact with the theory of phase transitions (see the discussion below), let us assume that the essence of the transition may be captured by a scalar order parameter $\varphi({\bf x}, t)$, equal to the fraction of the insulating phase in the sample \cite{Tapas, Limelette}. The time series of the sample resistance [Fig. 2 (a, b)] can then be approximately converted to the time series of the fluctuations in the spatial average of this order parameter $\delta \phi(t)$ using results from percolation theory \cite{percolation, SM}. Enhanced variance of the order parameter fluctuations $\langle\delta\phi^2(0)\rangle$ around the transition [Fig. 2 (e, f)] is the first indication of criticality \cite{CriticalOpalescence_Mott}.

Figure 3 (a, b) shows the corresponding normalized power spectral density (PSD) $S_R(f)/R^2$ in the frequency window $0.3-0.003$ Hz \cite{SM}  at different temperatures. The PSD obeys the characteristic $f^{-\mu}$ behavior \cite{AKR_NdNiO3, hartman_SlowingDownMott}, with an increase in the value to $\mu\approx 1.6$ around the transition, indicating a shift of the spectral weight of fluctuations to lower frequencies. Such slowing down has been previously observed in the resistance noise data at the Mott critical point \cite{hartman_SlowingDownMott} and simulations of the Ising model \cite{Chen-Yu}. Noise studies at the APT have also observed qualitatively similar features \cite{AKR_NdNiO3, Sambandamurthy_NdNiO3, Chandni} but the combination of the large hysteresis, a very sharp transition and the similarity in the behavior along the cooling and the heating runs in bulk V$_2$O$_3$ studied here has made the identification with the spinodal singularity more obvious. Our ongoing work on epitaxial NdNiO$_3$ thin films also qualitatively reproduces Fig. 3 (c, d), indicating its generality. Indeed recent works on a ferrite \cite{zhu} and VO$_2$ \cite{wang_schuller, del Valle} also invoke `pseudo-criticality' \cite{ikeda} to explain the slowed dynamics around APT.

\begin{figure}[t!]
\begin{center}
\includegraphics[scale=0.3]{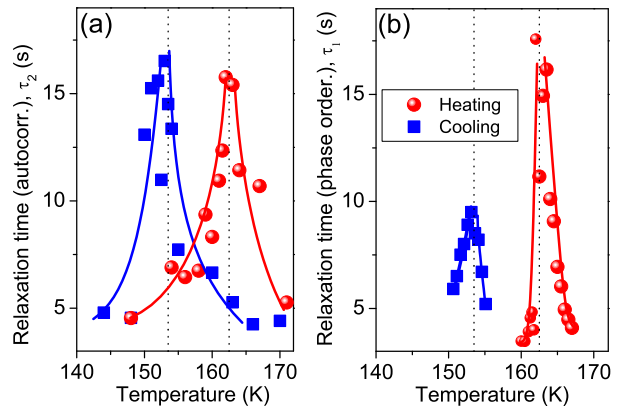}
\caption{ {\em Critical-slowing-down} is reflected in both $\tau_1$ and $\tau_2$ which, in the linear approximation, are equal to each other and $\tau_{k=0}$ and diverge with the susceptibility exponent: (a)  Temperature dependence of $\tau_2$ the relaxation time for the autocorrelation of the detrended resistance fluctuations,  and (b) $\tau_1$  the phase ordering time during heating and cooling.  See Supplemental Material for the details of the phase-ordering experiments. Solid lines are to guide the eye. The region between $\sim  153-162$ K is the metastable phase.
}
\end{center}
\end{figure}

In Fig. 4(a), this signature of slowing down is further evident from the relaxation time $\tau_2$ of the autocorrelation of the fluctuations \cite{SM}. In Fig. 4(b) we have also independently estimated the phase ordering time $\tau_1$ where the sample was shock-heated and quench-cooled respectively at $50$ K/min to the desired target temperature marked on the abscissa \cite{SM}. $\tau_2$ also shows a distinct enhancement around the transition \cite{SM}. Such an increase in the lifetime with deeper supersaturation (and consequent reduction in the energy barrier) within the metastable region would seem counterintuitive for a thermally-activated process like nucleation, but is the expected behavior at the spinodal \cite{binder_rpp}.

{\em The spinodal singularities.---}Let $F[\varphi]$ be the free energy functional of the system with an order parameter $\varphi({\bf x}, t)$ that denotes the fraction of the insulating phase in the material \cite{Tapas, Limelette}. The dynamical behavior this stochastic variable $\varphi({\bf x}, t)$, a non-conserved scalar, is then given by the dissipative Model A \cite{chaikin-lubensky}
\begin{equation}
{\partial\over \partial t} \varphi ({\bf x}, t)=-\lambda{\delta F[\varphi]\over \delta  \varphi ({\bf x})}  + \zeta({\bf x}, t).
\end{equation}
Here $\lambda$ is a kinetic parameter (that was previously determined to be $3.5$ s$^{-1}$ \cite{Tapas}) and  the thermal noise $\zeta({\bf x}, t)$ is as usual assumed to be $\delta$-correlated, i.e., $\langle \zeta({\bf x}, t)\zeta({\bf x^\prime}, t^\prime)\rangle=2\lambda k_BT \delta({\bf x}-{\bf x}^\prime)\delta(t-t^\prime)$, with zero mean \cite{chaikin-lubensky}. The angular brackets denote canonical ensemble averaging.

If $\varphi=\varphi_0$ for the metastable phase in the mean field approximation, we can Taylor-expand $F[\varphi]$ around this saddle point keeping only Gaussian fluctuations, viz. $F[\varphi]\approx F[\varphi_{0}]+ {1\over 2}\int d{\bf x} d{\bf x}^\prime \delta \varphi({\bf x})[\beta G({\bf x}- {\bf x}^\prime)]^{-1}\delta\varphi({\bf x}^\prime)$, where
$[\beta G({\bf x}-{\bf x}^\prime)] ^{-1}={\delta^2 F[\varphi_0]\over \delta \varphi({\bf x})\delta \varphi({\bf x}^\prime) } $ and $\beta^{-1}=k_BT$. $G({\bf x}-{\bf x}^\prime)$ is correlation function evaluated at $\varphi_0$ and let  $\tilde{G}_{\bf k}$ be its Fourier transform  \cite{chaikin-lubensky}. If $\delta \tilde{\varphi}_{\bf k}(t)$ is the Fourier transform of $\delta \varphi({\bf x},t)\equiv \varphi({\bf x},t)-\varphi_{0}$, the first two moments of $\delta \tilde{\varphi}_{\bf k}(t)$ evolve as
\begin{equation}
\langle\delta \tilde{\varphi_{\bf k}}(t)\rangle=\langle\delta \tilde{\varphi}_{\bf k}(0)\rangle\, \exp(-t/\tau_{\bf k}),
\end{equation}
\begin{equation}
\langle \delta\tilde{\varphi}_{\bf k}(t)\delta\tilde{\varphi}_{{\bf k}}(0)\rangle=V \tilde{G}_{\bf k}\, \exp(-t/\tau_{\bf k}).
\end{equation}
$\tau_{\bf k}=\beta \tilde{G}_{\bf k}/\lambda$ is the relaxation time for the ${\bf k}^{th}$ mode.

The sample resistance, after correcting for percolation effects \cite{SM, percolation}, tracks the spatial average of the order parameter \cite{Limelette}, i.e.,  essentially the zero-wave vector Fourier mode $\tilde{\varphi}_{k=0}(t)[=\int d{\bf x}\, \varphi({\bf x},t)]\equiv\phi(t)$. Thus our measurements  reflect  Eq. 2 and 3 with $k=0$. By the sum rule \cite{Goldenfeld}, the isothermal susceptibility  $\chi_{_T}=\beta\tilde{G}_{k=0}$. It follows that $\chi_T=[\delta^2F/ \delta \phi^2]^{-1}\rightarrow\infty$ at the spinodals, in strict analogy with its behavior at the critical point \cite{chaikin-lubensky}. Phase-ordering monitored through resistance should thus evolve as
\begin{equation}
\langle\delta \phi(t)\rangle=\langle\delta \phi(0)\rangle\, \exp(-t/\tau_0); \; \tau_0=\chi_T/\lambda,
\end{equation}
and the autocorrelation of the resistance noise should be related to
\begin{equation}
\langle \delta\phi(t)\delta\phi(0)\rangle=V k_BT \chi_T\, \exp(-t/\tau_0);\; \tau_0=\chi_T/\lambda.
\end{equation}
The phase ordering time $\tau_1$ and the autocorrelation time $\tau_2$ [Fig. 4] are equal to $\tau_0$ and diverge proportional to $\chi_{_T}$. Eq. 4 further shows that the variance $\langle \delta{\phi}^2(0)\rangle$ should also diverge as $\chi_T$ (`critical opalescence'), in agreement with Fig. 2 (e, f) \cite{Sasaki}. Regarding the slopes of the PSD [Fig. 3], the linear theory would predict Brownian dynamics ($\mu=2$) at the spinodals. Theoretical estimates for the critical point of the two-dimensional Ising model under Glauber dynamics suggest $\mu \sim 1.8$ \cite{Chen-Yu}.

{\em Discussion.---}
It had long been an open question whether the spinodal singularity can be seen in finite temperature experiments. While for short-ranged interactions, the spinodals are physically meaningless \cite{binder_rpp}, in the opposite limit of infinite-ranged interactions, the mean field theory becomes exact \cite{Penrose-Lebowitz}. Thus the point of contention is whether the spinodal may be a meaningful concept for long but finite-ranged interactions \cite{FN:Spinodal_lightScattering}. Theoretically, the issue is addressed by constructing, as usual \cite{chaikin-lubensky}, a relationship (Ginzburg criterion) between the relative order parameter fluctuations, system dimensionality, and the range of interactions as one approaches the spinodal singularity  \cite{binder_rpp, Gulbahce, Unger-Klein}. Provided the interactions are sufficiently  long-ranged, one can get close enough to the instability such that finite temperature effects broaden the transition but may not completely mask it \cite{Gulbahce, Stephanov, Klein_fluctuations}, mimicking a finite-size effect. This is indeed what we see; the singularity still manifests as a very discernable (but finite) growth in the susceptibility. That we have long-range interactions in the system \cite{Mukamel} is independently indicated by the observations of broken ergodicity [Fig. 1 (b)] and very slow relaxation [Fig. 4], which are otherwise unusual for hard condensed matter systems. Furthermore the robust presence of hysteresis itself indicates long-range forces on rigorous theoretical grounds \cite{sewell}.

The electronic phase transitions in a number of correlated electron systems are often also accompanied by abrupt structural transitions on account of the large polaronic coupling \cite{Peil, Kalcheim}.
The resulting strain fields are the source of long-range interactions \cite{Klein_Nucleation_elastic, Klein_fluctuations, Miyashita_PRB2009}. Indeed a structural transition does seem like the feature shared across disparate systems undergoing APT with pronounced hysteresis \cite{McWhan_V2O3, Tapas, mcleod, Kalcheim, AKR_VO2, liu, wang_schuller, del Valle, Peil, Sambandamurthy_NdNiO3, AKR_NdNiO3, Post, Manganite_singh, Chandni, Chaddah, Ong_CDW, zhu, Kahn_Martinez}. These are all candidates for the zeroth-order transition belonging to the mean-field `universality class' \cite{Mori-Miyashita-Rikvold, Miyashita-Konishi-Nishino-Tokoro-Rikvold}.

We have not discussed disorder. Though the phase coexistence and ramified fractal-like structures in the transition regions \cite{liu} are also explained by the non-classical nucleation expected at the spinodal \cite{klein-monette, Klein_fluctuations} without invoking disorder, it is perhaps essential for understanding avalanches \cite{Nandi, sethna} or even the return-point memory \cite{sethna, Keim_RMP} seen in Fig. 1 (b). While one would naturally expect the smooth evolution suggested by Eq. 1 to be not quite valid \cite{FN:annealed-disorder}, the spinodal singularity itself is expected to be robust to disorder for sufficiently long range interactions \cite{Liu-Klein}. But this is clearly an issue that requires further systematic investigation.

Finally, catastrophes \cite{Gilmore} in natural and social systems---earthquakes, ecosystem collapse, climate change, onset of depression, epilepsy, market crash---are sometimes modeled after such zeroth-order transitions, with the spinodal singularity signifying the loss of `resilience' \cite{Scheffer-Review}. Anticipating them is of course an important objective of complex systems science. The idea that fluctuations can carry precursory signatures of such an impending event, in the form of critical-slowing down accompanied by an enhanced variance of fluctuations \cite{Scheffer-Review}, is perfectly vindicated by our work.

It is a pleasure to thank Arup Raychaudhuri for critical advise on the noise measurements, Amit Ghosal for comments on the manuscript, and Satyabrata Raj for technical advice on sample preparation.   

\begin{center}
{\large {\bf Supplemental Material}}
\end{center}
In this Supplemental Material we discuss the details of the resistance noise experiments including a discussion on the temperature stability during the time series measurements and the method employed for detrending the time series. Raw data of the autocorrelation of the resistance fluctuations [Fig. 4 (a) (Main text)] and of the phase-ordering experiments [Fig. 4 (b) (Main text)] are discussed. The algorithm to convert the resistance noise data to the order parameter fluctuation using the approximate percolation model [Fig. 2 (e, f) (Main text)] is discussed. Finally we show that a rough but independent thermal analysis measurement also indicates  sharp maxima in the phase ordering times around the spinodals.
\tableofcontents
\section{Sample}
Single phase polycrystalline V$_2$O$_3$ samples were used in the measurements. The details can be found in the Supplementary Material of our previous work \cite{Tapas}. As already mentioned, the samples were thermally cycled hundreds of times such the measurements were reproducible.

\section{Noise Measurements}
Transition metal oxides like V$_2$O$_3$, VO$_2$, or NdNiO$_3$ are known to have a large Hooge parameter \cite{AKR_NdNiO3}. A conventional 4-probe geometry is thus sufficient and the experimental constraints on reducing the background noise are less stringent. The measurements nevertheless require extreme caution and multiple crosschecks. Unlike semiconductors, it can sometimes be hard to get reproducible data and the samples require a fair amount of training.

Previous work on noise measurements around the phase transition have often involved ramping the temperature at a given rate ($\sim 0.1$-$10$ K/min) and constantly recording the resistance during the temperature ramp \cite{{hartman_SlowingDownMott}, Chandni}. This is them broken into segments within temperature windows where the time series is detrended and analyzed. While we attempted to do this, even for the lowest temperature ramp rate of $0.1$ K/min (below which the measurements become untenably long), the change in the resistance around the transition was too abrupt and it was difficult to reliably detrend the time series. Therefore the time series were recorded with the temperature kept fixed a given value.

\subsection{Experimental setup}
The setup consists of a liquid nitrogen-cooled cryostat with a temperature range from $77$-$300$ K. Gold wire contacts made with silver epoxy paste in van der Pauw geometry were used through the study.

Given that the characteristic relaxation time, previously inferred from the time-dependent Landau dynamics \cite{Tapas}, is about $3.5$ s$^{-1}$, we focussed on much lower frequencies ($<1$ Hz) than are typically used ($<10-100$ Hz). This also proved helpful because the sample resistance could be $>10^5\,\Omega$ in the insulating phase and capacitive effects begin to be prominent at higher frequencies. Note that it is necessary that the lockin excitation frequency be about $50-100$ times larger than the high frequency cut-off in the power spectral density to avoid any spurious settling time effects associated with the lowpass filter required for the homodyne detection, a safe settling time being $>5$, say $10$, lockin time constants.

The measurement circuit is shown in Fig. 1 (supplement). The measurements involved exciting the sample with an oscillatory voltage at the frequency of about $23$ Hz, with a large $1$ M$\Omega$ low noise reference resistance connected in series. The voltage across the other two contacts was fed into a high input impedance (100 M$\Omega$) voltage preamplifier (Stanford Research Systems SR560) equipped with bandpass filters which also helped in cutting down the aliasing effects. The output of the preamplifier was fed into the lockin amplifier (Stanford Research Systems SR830) whose time constant was chosen to be $300$ milliseconds. The output of the lockin amplifier was read out the rate of $10$ Hz (time stamp from the computer clock was made on each data point). For the purpose of analysis of the power spectral density (PSD) of the time series, the upper frequency cut off was chosen to be $0.3$ Hz (i.e, a very safe $10$ time constants). The sample current was also independently monitored by another locking amplifier (Stanford Research Systems SR830). The sample temperature was controlled with a proportional-integral-derivative (PID) based digital temperature controller (Lakeshore 336) with a temperature resolution of 1 millikelvin. The measurement electronics was electrically isolated from the measurement computer and proper shielding and grounding of the set up was ensured.

\renewcommand\thefigure{1 (supplement)}
\begin{figure}[h!]
	\includegraphics[scale=0.2]{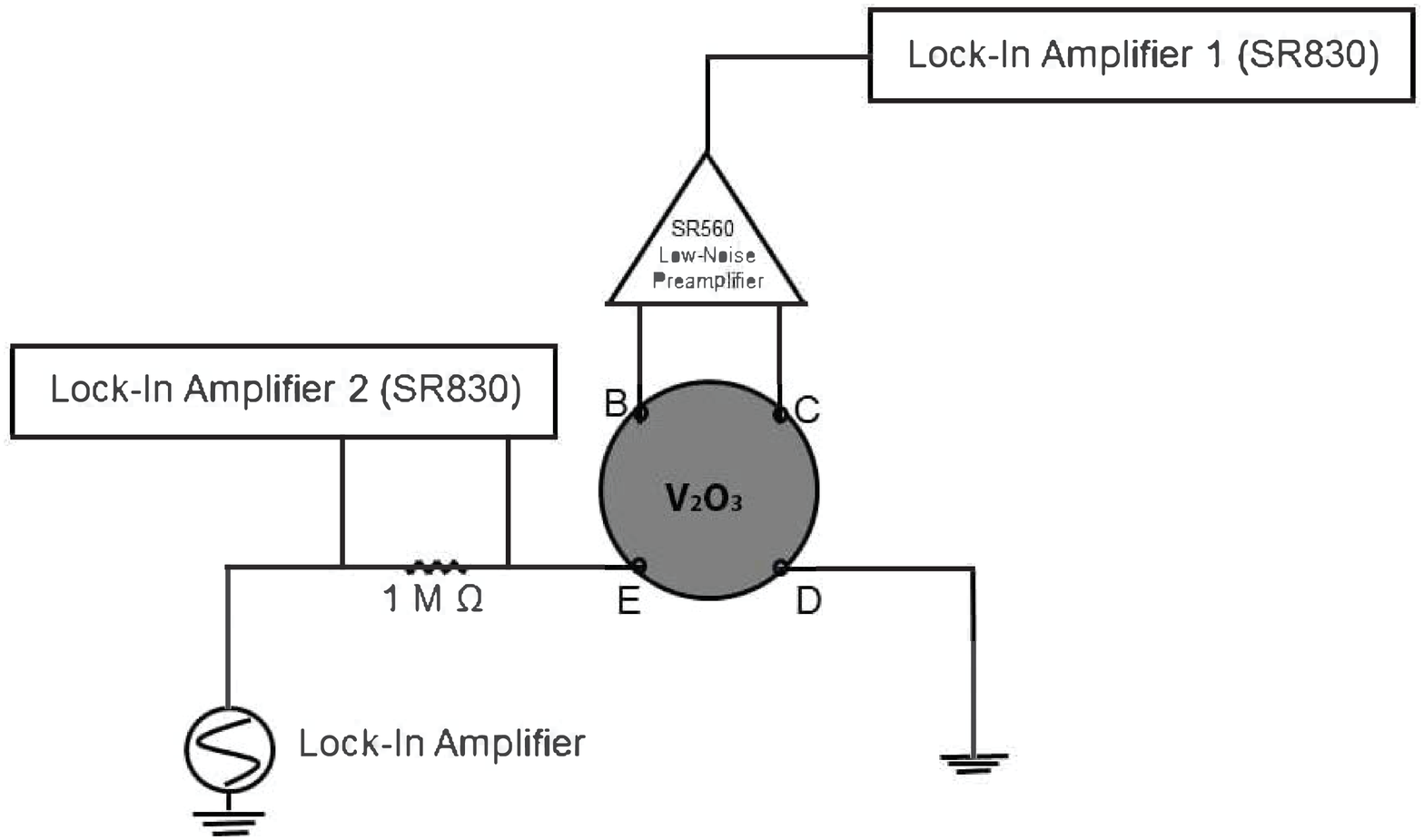}
	\caption{Circuit for the resistance noise measurements.}
	\label{Measurement_setup}
\end{figure}

\subsection{Temperature stability}
As the sample response in the metastable phase is highly history dependent, the sample was always cooled (heated) to well outside the metastable phase before each time series measurement and then the temperature was ramped up (down) to the desired set point. Then the voltage fluctuation of the sample with time were recorded. For each temperature set point, the best PID control parameter values were determined. Typically, the voltage time series was recorded for many hours and then an appropriate segment (of duration $>60$ minutes) was picked out by manual inspection, where the fluctuation in the temperature was less than $10$ mK, usually better and limited by the digitization noise [Fig. 2 (supplement)].

\renewcommand\thefigure{2 (supplement)}
\begin{figure}[h!]
	\includegraphics[scale=0.3]{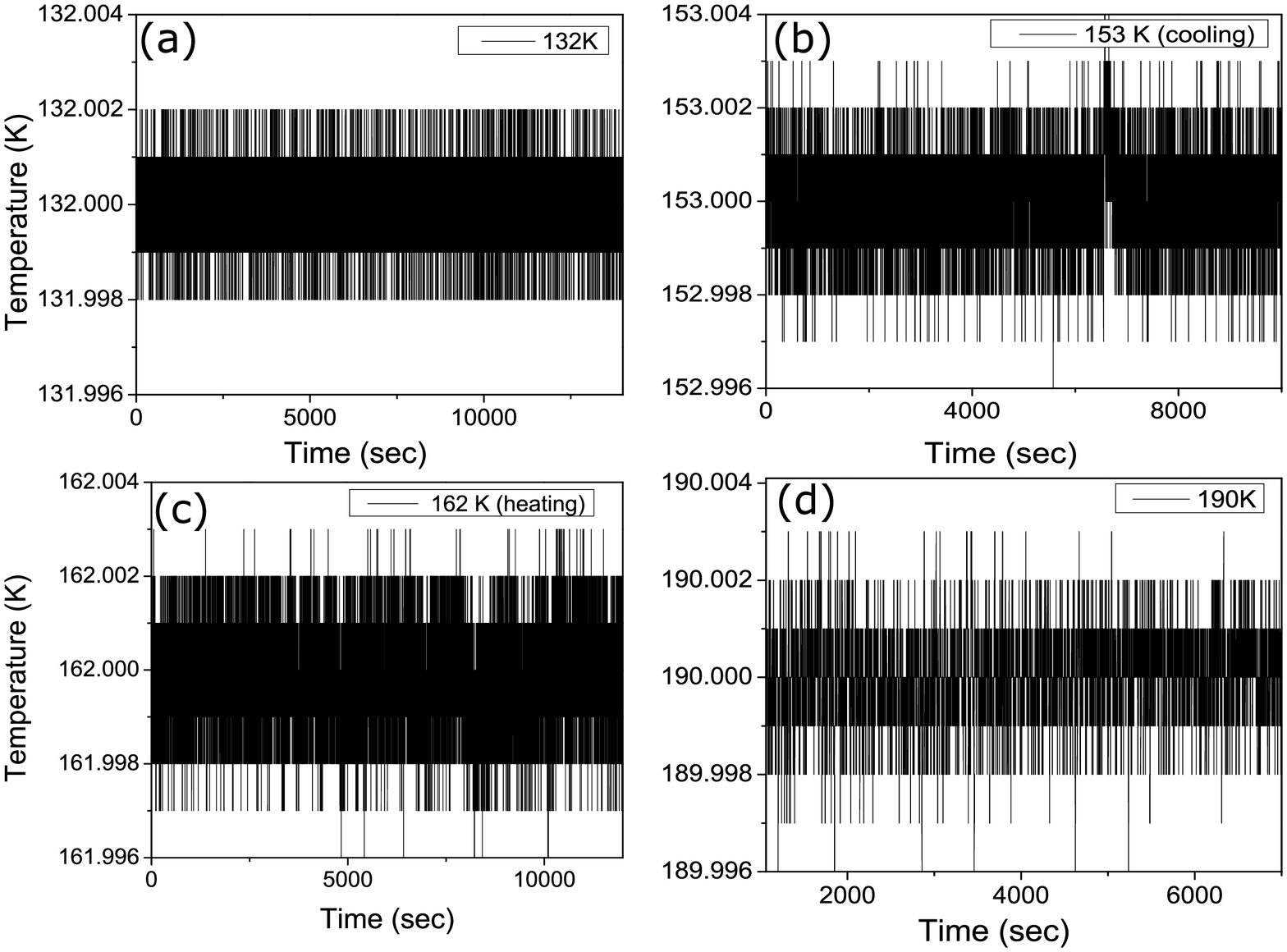}
	\caption{Some typical temperature profiles in the time windows where the resistance time series data were analyzed. (a) Insulating region (132 K), (b) cooling spinodal (153K), (c) heating spinodal (162K), and (d) metallic region (190K). Note that the temperature stability is essentially limited by the resolution set by digitization.}
\end{figure}

\section{Analysis of the voltage time series}
On reaching the set point (especially around the spinodal regions), the sample displays phase ordering where a substantial change in the resistance occurs within a few tens seconds. The temperature also stabilizes during this time. The lockin amplifier was then offset to zero and the gain enhanced to get the best possible voltage resolution. As mentioned above, from the many hours long time series, we only take the part where the temperature is most stable.
It was cross-checked that even at the spinodals (where the change of the resistance with temperature is the sharpest), the possible resistance fluctuations due to the temperature fluctuations were at least one order of magnitude smaller than the measured noise. To ensure that the measured noise was (i) indeed from the sample and not the associated electronics and (ii) the bias on the sample was small enough to not cause heating, we verified that the power spectral density remain unchanged (i.e., the power spectrum scaled proportional to $\langle V^2\rangle$) when the bias across the sample was doubled.

\renewcommand\thefigure{3 (supplement)}

\begin{figure}[h!]
	\includegraphics[scale=0.3]{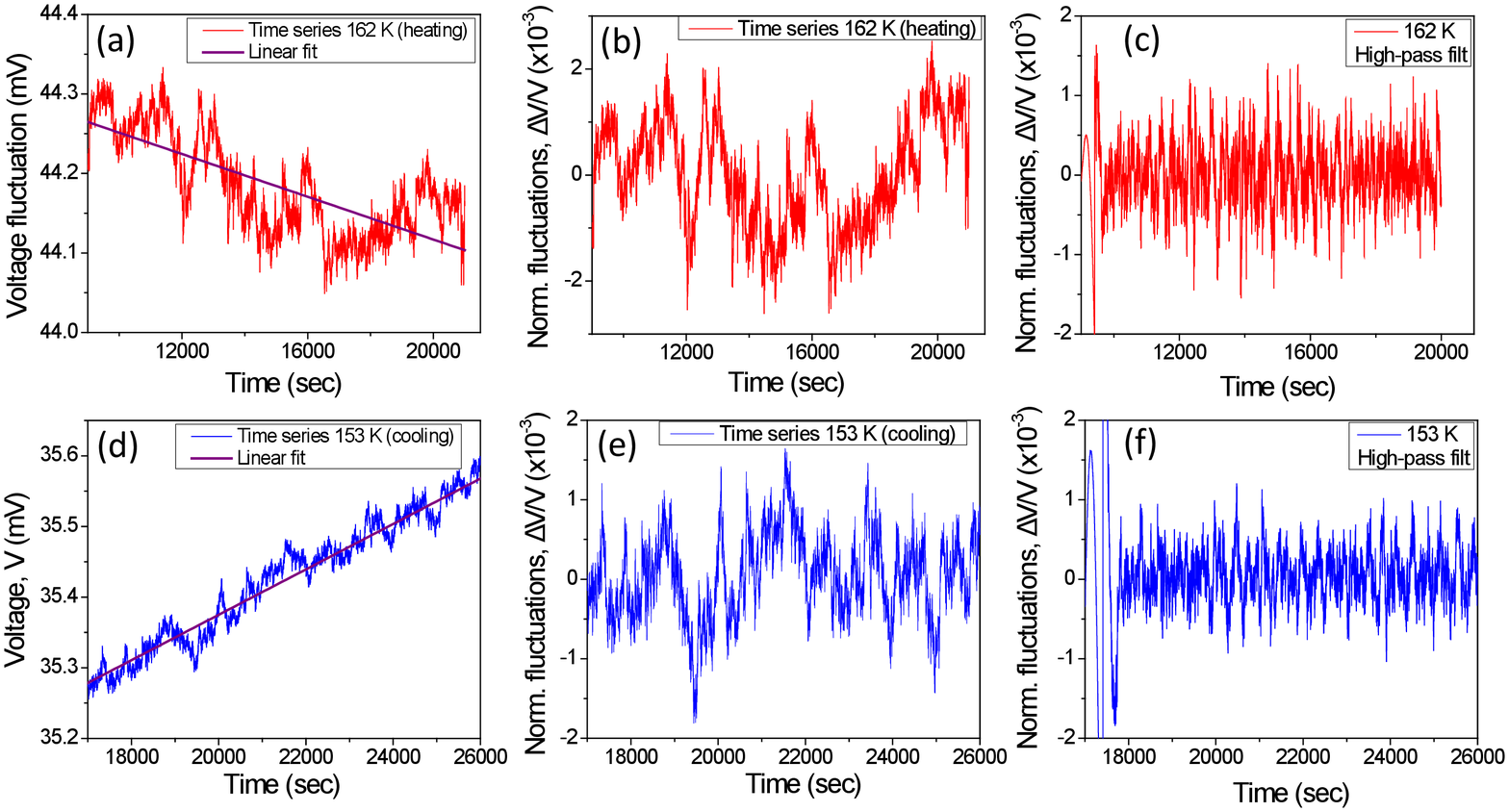}
	\caption{(a) The raw time series of the voltage fluctuations at 162 K, which is close to the heating spinodal. Note that there is a small but significant ($< 1\%$ in about $10^5$ seconds) monotonic drift due to the slow phase ordering. Solid line shows a straight line fit. (b) Detrending by subtracting this straight line. This detrended time series is still dominated by large low frequency fluctuations. These are real but it is hard to get good statistics on these. As a result, the assessment of the autocorrelation, variance and the probability distribution function of the fluctuations is not reliable. (c) A simple alternative is to directly highpass filter the data in shown in (a). This removes both the linear trend and puts a systematic low frequency cut-off on the bandwidth too. Note that filtering has the (well-known) artifact of introducing spurious oscillations in the beginning of the time series. This part of the time series is ignored. (d, e, f) The same for the cooling spinodal time series at $153$ K. }
\end{figure}

Around the spinodals, as the phase ordering is a slow process, there is always a small monotonic drift (about $<1\%$ in $>3$ hours) in the value of the sample resistance [Fig. 3 (a, d) (supplement)]. The time series thus need to be detrended.

The second problem, again increased at the spinodals due to the slowing down, is that there is a large spectral weight in the very low frequency components of the power spectral density (PSD), which can be seen as large fluctuations occurring on the time scale of tens of minutes and modulate the whole time series. While these are real, clearly it is not possible to get sufficient statistics on such large time scales. While the determination of PSD was robust to the method of detrending (as the lowest frequencies which are most affected by detrending could be thrown away), quantities like the variance, autocorrelations and the probability density functions of the fluctuations are very hard to determine precisely.

A simple and robust solution was found in passing the raw recorded time series through a digital high pass filter and limiting the low frequency to about $3$ millihertz. This was implemented using the {\tt{designfilt}} function in MATLAB software, with the following parameters: Stopband frequency: $0.0008$ Hz;  Passband frequency: $0.003$ Hz, Stopband attenuation: 50, Passband ripple: 1 and Sample rate: 10 Hz.

\renewcommand\thefigure{4 (supplement)}
\begin{figure}[h!]
	\includegraphics[scale=0.28]{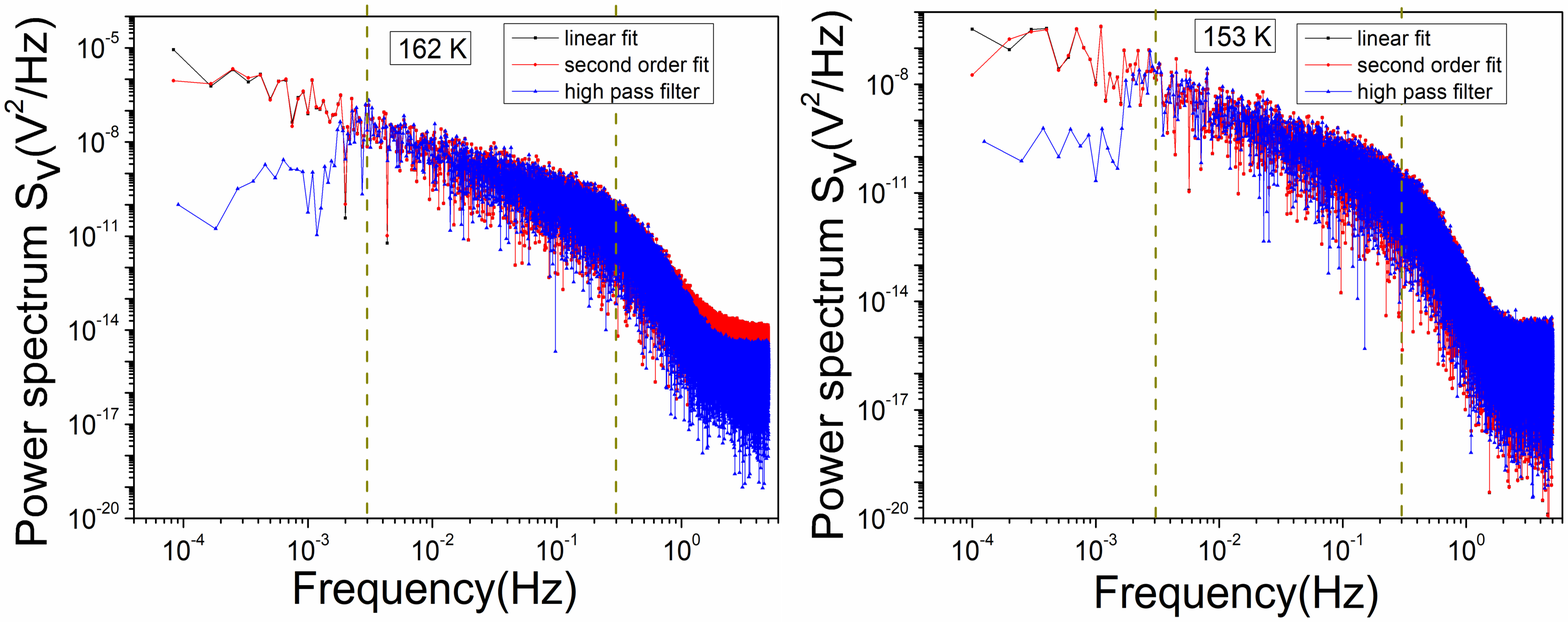}
	\caption{Power spectrum of the voltage time series (a) around the cooling spinodal at $153$ K and (b) close to the heating spinodal $162$ K calculated by three different methods of detrending, namely via a linear subtraction, by subtracted a best fit quadratic and via high pass filtering. In case of linear and quadratic fits, the difference is only at the lowest frequency. We have band-width limited the measurement to a lower frequency of $0.003$ Hz. The high frequency cut-off ($0.3$ Hz) is imposed by the low pass filter of the lockin amplifier. }
\end{figure}

\section{Robustness of PSD slope $\mu$} Fig. 4 (supplement) and Fig. 5 (supplement) shows that the inferred slope $\mu$, $S_V/\langle V\rangle^2\sim f^{-\mu}$ of the PSD between $0.3-0.003$ Hz is essentially robust to the method used for detrending the time series. Since the number of data points in a frequency octave vary drastically from octave to octave, a naive least square fitting (on the log-log scale) to determine $\mu$ puts a disproportionate weight on the high frequency region of the PSD. To circumvent this problem, the PSD were smoothed and the data between $0.003-0.3$ Hz were interpolated, equispaced on the log scale. These are the PSDs shown in Fig. 2 (a, b) (Main text).

\renewcommand\thefigure{5 (supplement)}
\begin{figure}[h!]
	\includegraphics[scale=0.3]{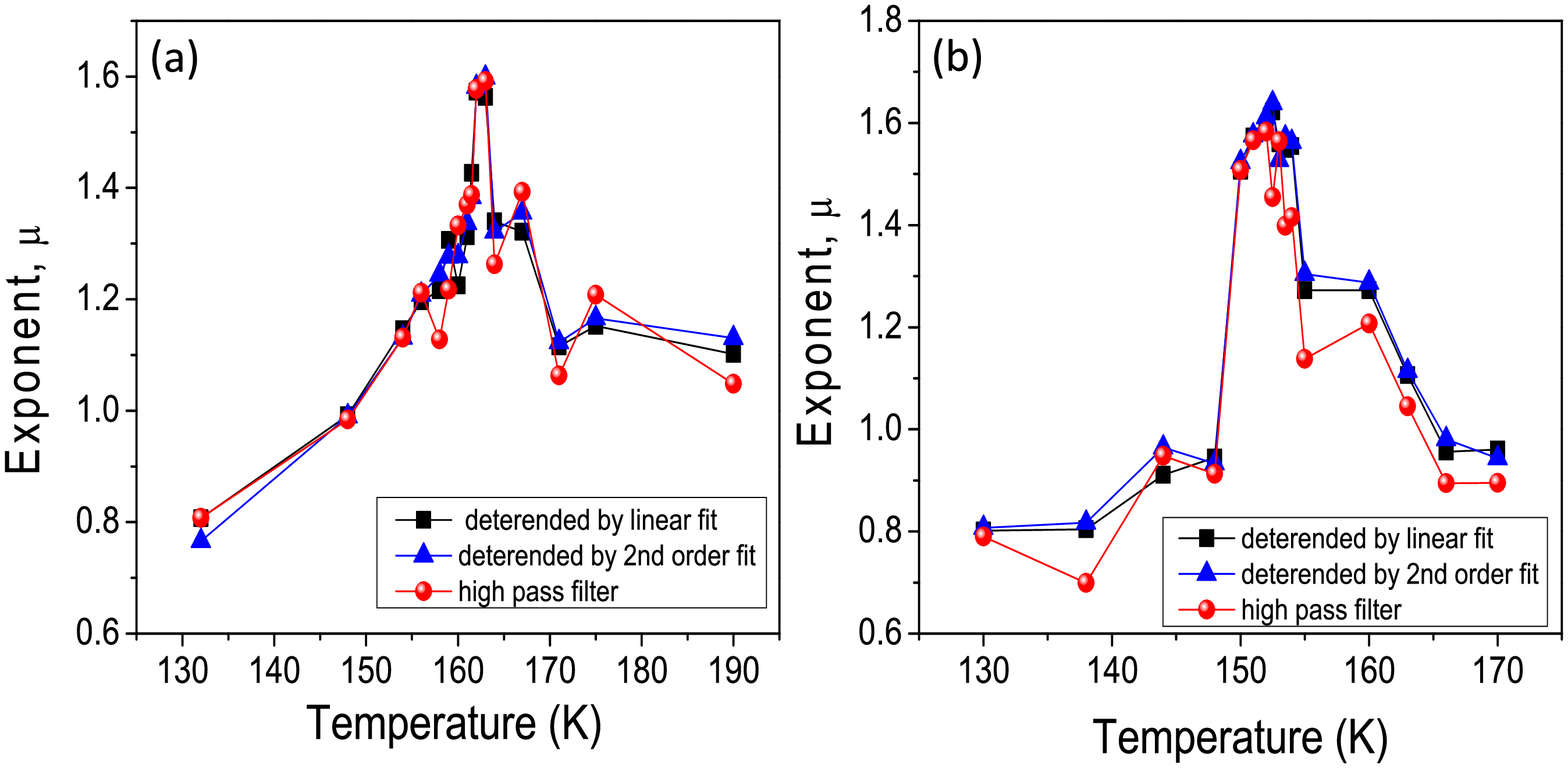}
	\caption{Inferred values of the noise PSD exponent $\mu$ for the time series detrended using a linear subtraction, substraction of the best fitting quadratic, and high pass filtering at different temperatures for the measurements done under (a) heating and (b) cooling protocols showing that the conclusions of the slowing down are robust to the method of detrending.}
\end{figure}

\section{Order parameter from the resistance}
It is interesting to attempt to convert the sample's resistance to a more microscopic and theoretically interesting quantity, the order parameter. As is argued in our previous work \cite{Tapas} and also in previous work on V$_2$O$_3$ around the critical point \cite{Limelette}, the fraction of the insulating phase in the material could reasonably be taken as the order parameter. But the relationship between the measured resistance and the insulating fraction is a monotonic but a complicated non-linear function. We shall use the equation given by McLachlan, which is based on a general effective medium (GEM) theory for percolative transitions.
If $\sigma_E$ is the effective electrical conductivity of a binary metal-insulator mixture, the GEM equation is given by \cite{percolation}
$$(1-f)\frac{\sigma_I^{1/t}-\sigma_E^{1/t}}{\sigma_I^{1/t}+A\sigma_E^{1/t}}+f\frac{\sigma_M^{1/t}-\sigma_E^{1/t}}{\sigma_M^{1/t}+A\sigma_E^{1/t}}=0$$
where f is the volume fraction of the metallic phases and
$A = (1-f_c)/ f_c$, $f_c$ being the volume fraction of metallic
phases at the percolation threshold, and t is a critical exponent
which is close to 2 in three dimensions.The constant
$f_c$ depends on the lattice dimensionality, and for 3D its value
is $0.16$.

Here we are considering the insulator fraction ($\varphi({\bf x},t)$) as order parameter and the conductivity is related to the spatial average of $\int\varphi d^x\sim \phi(t)$. Hence the relationship between $\phi$ and the $f$ defined above is simply $\phi = 1-f$. By using the relation between conductivity and resistance ($\sigma= \frac{l}{AR}$ where $l$ and $A$ are constants) we can rewrite the equation in terms of the sample resistance, viz.

\renewcommand\theequation{S1}
\begin{equation}
\varphi\frac{R_I^{-1/t}-R_E^{-1/t}}{R_I^{-1/t}+AR_E^{-1/t}}+(1-\varphi)\frac{R_M^{-1/t}-R_E^{-1/t}}{R_M^{-1/t}+AR_E^{-1/t}}=0
\end{equation}

The data in Fig. 8 (supplement) are inferred using the above equation.

More interestingly, using this relation we can approximately convert resistance fluctuations to the fluctuations of the order parameter, plotted Fig. 2 (e, f) (Main text). The result is in accordance with the theoretical expectation  [Eq. 3 (Main text) with $k=0$ and $t=0$] of observing a `divergence' at the two spinodals.

\section{Autocorrelation of fluctuations}
In Fig 4 (a) (main text), we have plotted the relaxation time $\tau_2$ for the autocorrelation of fluctuations. $\tau_2$ at different temperatures were inferred from the numerically calculated autocorrelation function $A(t)$ of the detrended time series, where $A(t)$ is defined as
\begin{equation}
A(t)\equiv\langle \delta R(t^\prime)\cdot\delta R(t+t^\prime)\rangle_{t^\prime}.
\end{equation}
$\langle \cdot\cdot\cdot\rangle_{t^\prime}$ denotes the time average. Figure 6 (supplement) shows $A(t)$  at a few temperatures. It can clearly be seen that the decay is significantly slower around the transition regions. Note that since the measurements were done with a lockin time constant of $300$ milliseconds, we would expect to have a low-pass-filter-circuit-induced correlations which may persist up to a few ($\lesssim 5$) time constants, which is about $1-2$ seconds.
\renewcommand\thefigure{6 (supplement)}
\begin{figure}[h!]
	\includegraphics[scale=0.28]{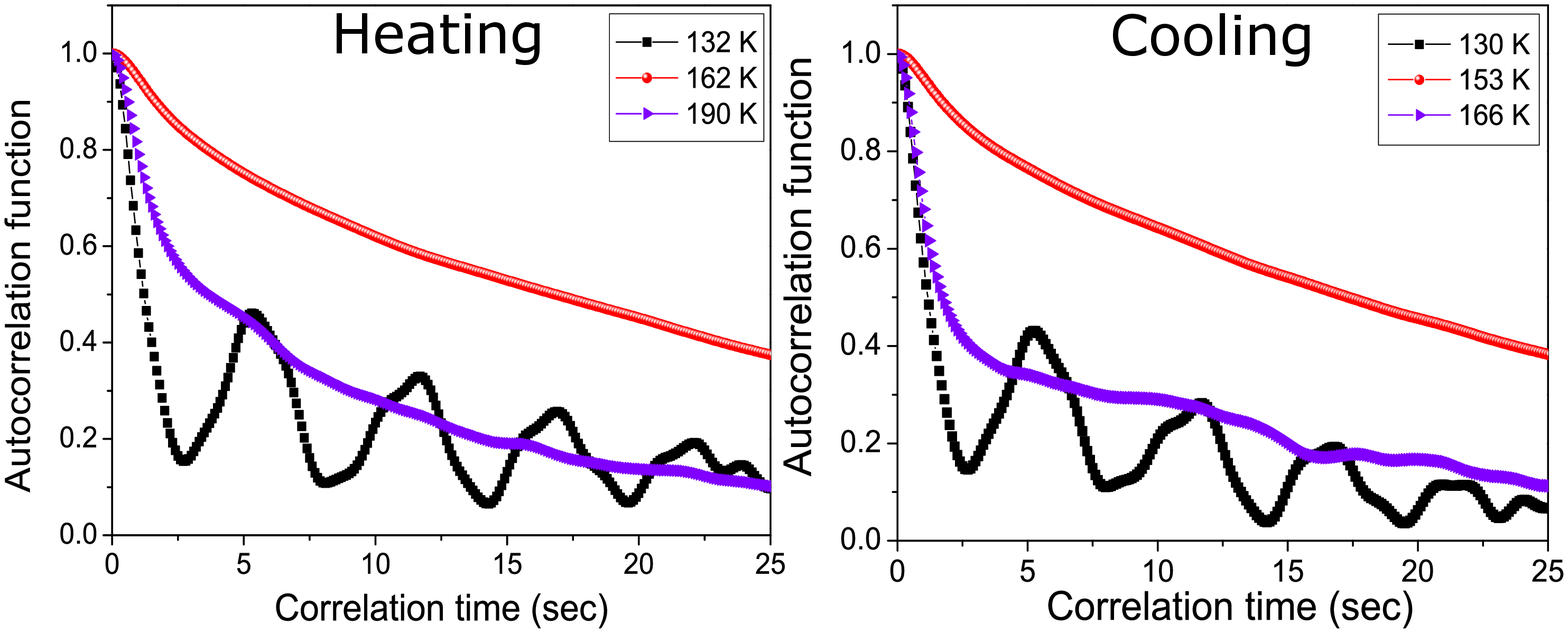}
	\caption{Autocorrelation. The reason for the oscillatory behavior of the autocorrelation in the insulating phase is not understood. Note however that the envelope shows a much faster decay compared to the spinodals. }
\end{figure}
The relaxation time $\tau_2$ is the characteristic decay time of the autocorrelation $A(t)$ and there is some arbitrariness in the definition unless one forces a model. It is only within the {\em linear theory} that the autocorrelation function is expected to be of a simple exponential form $[$Eq. 5 (main text)$]$, namely
\begin{equation}
A(t)\sim \exp[-t/\tau_0]\; \;\; \;\; \; \textrm{(linear \, theory\, only)}.
\end{equation}
and the real $A(t)$ never fits this. Luckily, we are not interested in the exact numerical value but only wish to (unambiguously) show that the characteristic scale of the relaxation time rapidly grows around the spinodals. This fact is not sensitive to the details of the definition.

\renewcommand\thefigure{7 (supplement)}
\begin{figure}[!h]
	\begin{center}
		\includegraphics[scale=0.32]{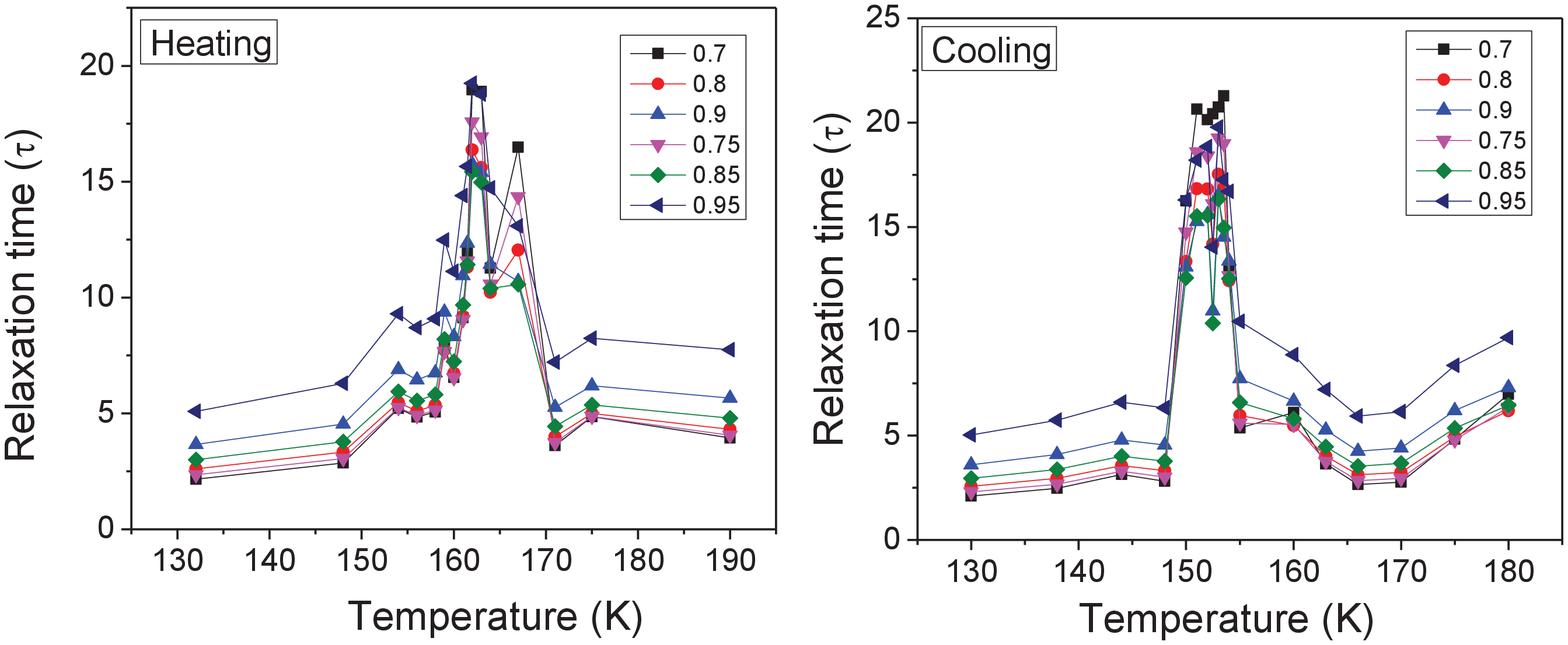}
		\caption{The characteristic autocorrelation time scale estimated for different values of  $\kappa$ shows the same characteristic trend. For a perfect monoexponential decay, all the curves would have coincided.}
	\end{center}
\end{figure}
But one can always extract a characteristic decay time $\tau_2$ assuming such a monoexponential decay and by finding out the value of the time $t_1$ when the the autocorrelation function has a value $\kappa$ with $0<\kappa<1$, e.g. if one chooses $\kappa=0.8$, $0.8=e^{-t_1/\tau_2}$. So by reading off the time $t_1$ from the autocorrelation graph, one can estimate $\tau$
\begin{equation}
\tau_2= \frac{t_1}{\ln(1/\kappa)}
\end{equation}
While there is an arbitrariness about which value of  $\kappa$ to take, it was reassuring to find that the trends in the temperature dependence of the characteristic time  $\tau$ and the observation of its enhancement around the transition does not depend on the value of $\kappa$ chosen (up to about $\kappa=0.7$ below which $A(t)$ in the insulating phase begins to oscillate with a decaying envelope). We considered $\kappa=0.9$ for Fig 4 (main text).
\section{Phase-ordering time}
In addition to the measurement of the relaxation times of the autocorrelation of the fluctuations, Fig. 4 (a) (Main text) also shows the characteristic phase ordering time $\tau_1$.
Figure 8 (supplement) shows the data used to infer $\tau_1$ at different temperatures around the two spinodals. The measurements involved shock-heating (cooling) the sample to a given set point at the rate of 50 K/min, starting deep into the insulating (metallic) phase. On reaching the set-point, the temperature was quickly stabilized and the change in the resistance with time was recorded. The resistance data were then converted to the order parameter using the transformation discussed above and the phase ordering time $\tau_1$ was inferred by fitting the equation
\renewcommand\theequation{S2a}
\begin{equation}
\varphi(t)=[1-\varphi_\infty]\exp(-t/\tau_1) + \varphi_\infty
\end{equation}
to the observed order parameter evolution after the shock heating quench and 
\renewcommand\theequation{S2b}
\begin{equation}
\varphi(t)=\varphi_\infty [1-\exp(-t/\tau_1)]
\end{equation}
for the evolution after the cooling quench [Fig. \ref{Fig8_Supp}]. 

\renewcommand\thefigure{8 (supplement)}
\begin{figure}[h!]\label{Fig8_Supp}
	\includegraphics[scale=0.3]{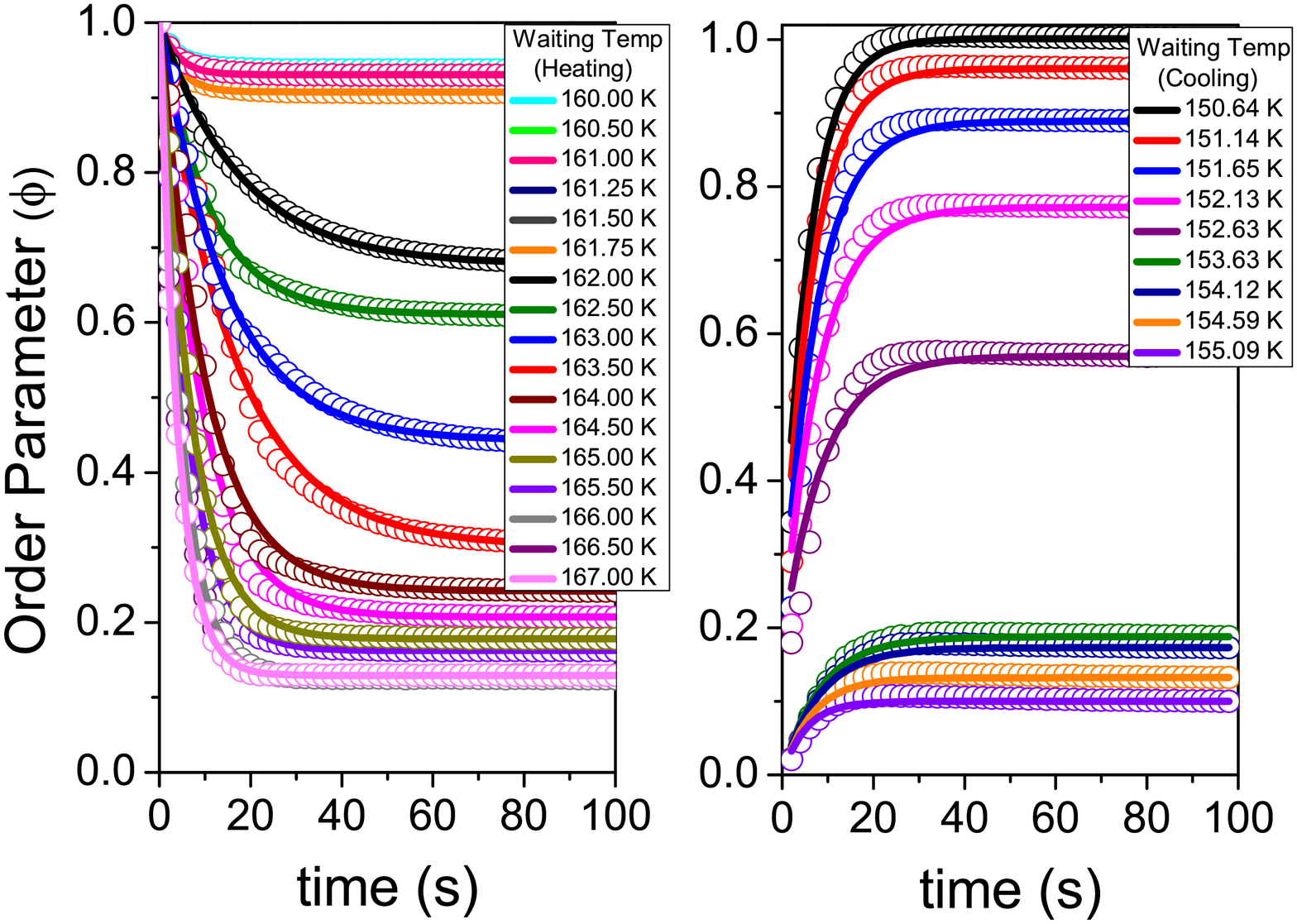}
	\caption{Phase ordering after a shock-heating (left) and cooling (right) quench. Solid lines are fits to Eq. (20).}\label{Fig8_Supp}
\end{figure}

\section{Time constant measurement using DTA technique}
In this section, we will show that the observed slowing down can also be qualitatively inferred from a completely independent thermodynamic measurement using differential thermal analysis (DTA). While this novel technique has some limitations and is only qualitative, it has the advantage that the extent of phase ordering is directly estimated from the amount of the latent heat measured. The method is an extension of the well-known first-order-reversal-curve (FORC) measurements of the hysteresis loops. The details of the DTA apparatus can be found in Ref. \cite{Tapas}.

\renewcommand\thefigure{9 (supplement)}
\begin{figure}[h!]\label{FORC}
	\includegraphics[scale=0.34]{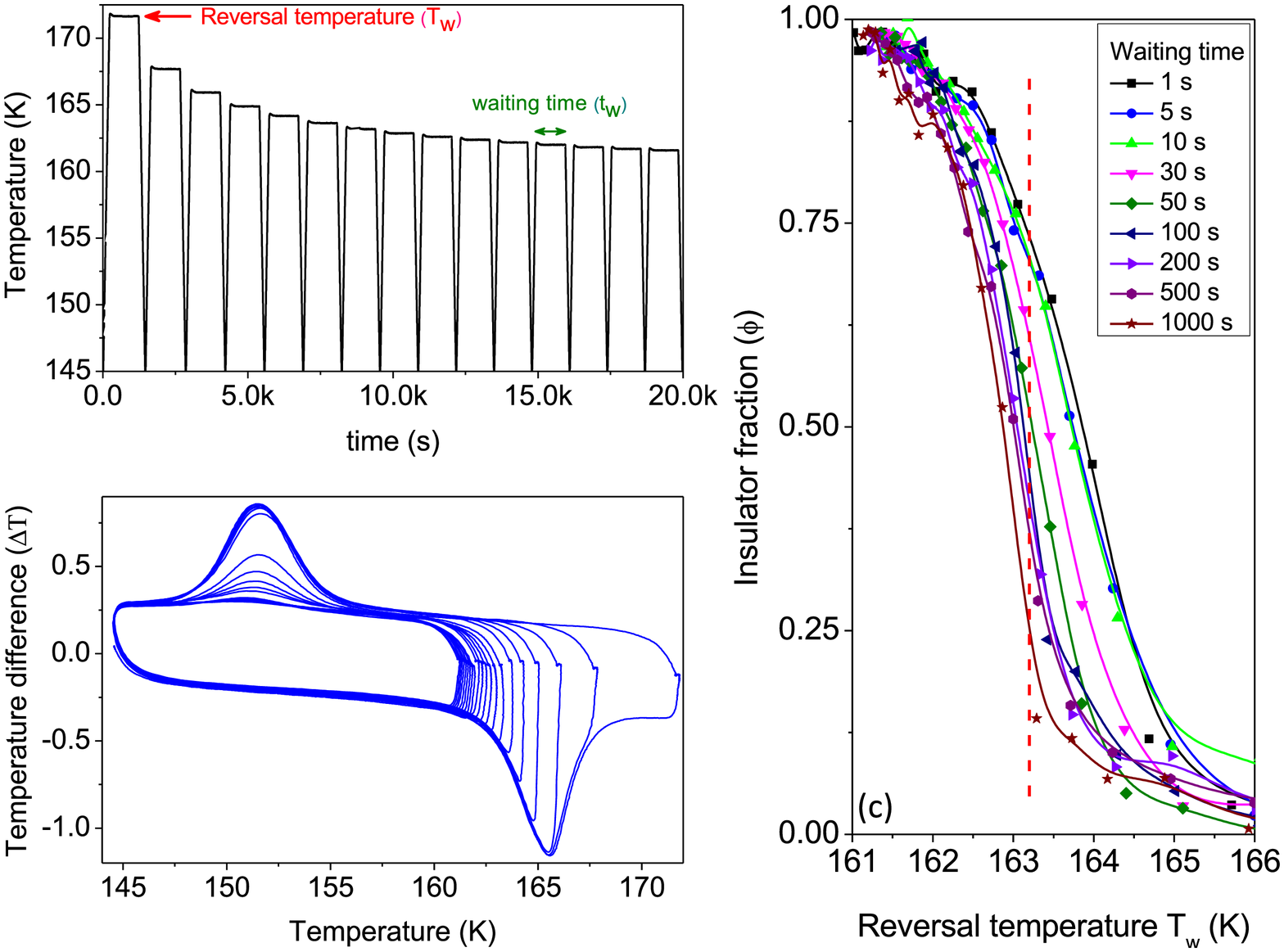}
	\caption{(a) Temperature profile for the heating FORC measurement. The waiting time was fixed (1000 s) for this set of measurements (Left-Right arrow). The Leftwards arrow corresponds to the first reversal temperature (T$_w$). (b) DTA signal for different wait temperatures. The area of the DTA peak at $T \sim 152 $K is a measure of the amount of material transformed due to the quench and wait at reversal temperature (T$_w$). (c) The insulator (phase) fraction versus reversal temperature (T$_w$) for different waiting time.}
	\label{FORC}
\end{figure}

The experiment involves, starting from a temperature $T_s$ (say) deep in the insulating phase and then shock-heating the sample at the rate of 50 K/minute to a given target temperature T$_w$ where T$_w$'s are around the heating spinodal ($\sim 162$ K). Then the sample temperature is kept fixed at this T$_w$ for a given wait time t$_w$, after which the sample is rapidly quench-cooled at the rate of 50 K/minute to the starting temperature $T_s$. As the sample is cooled, it will cross the cooling spinodal ($\sim 152$ K) and latent heat will be released. The main idea of the experiment is that this latent heat is proportional to the extent of the phase ordering corresponding to (T$_w$, t$_w$). Figure 9 (a) (supplement) shows the temperature protocol for a given wait time t$_w$ and Fig. 9 (b) (supplement) shows that different amount of latent heats are measurement depending on T$_w$. The complete measurement of the phase ordering [Fig. 9 (c) (supplement)] involves repeating this for different wait times, $\lbrace t_w=1, 5, 10, 30, 50, 100, 200, 500, 1000\rbrace$ seconds. t$_w=1000$ s corresponds to (almost) complete phase transformation and would yield the temperature dependence of the order parameter under quasistatic conditions, as was already discussed previously \cite{Tapas}.

\renewcommand\thefigure{10 (supplement)}
\begin{figure}[h!]\label{time_const}
	\includegraphics[scale=0.32]{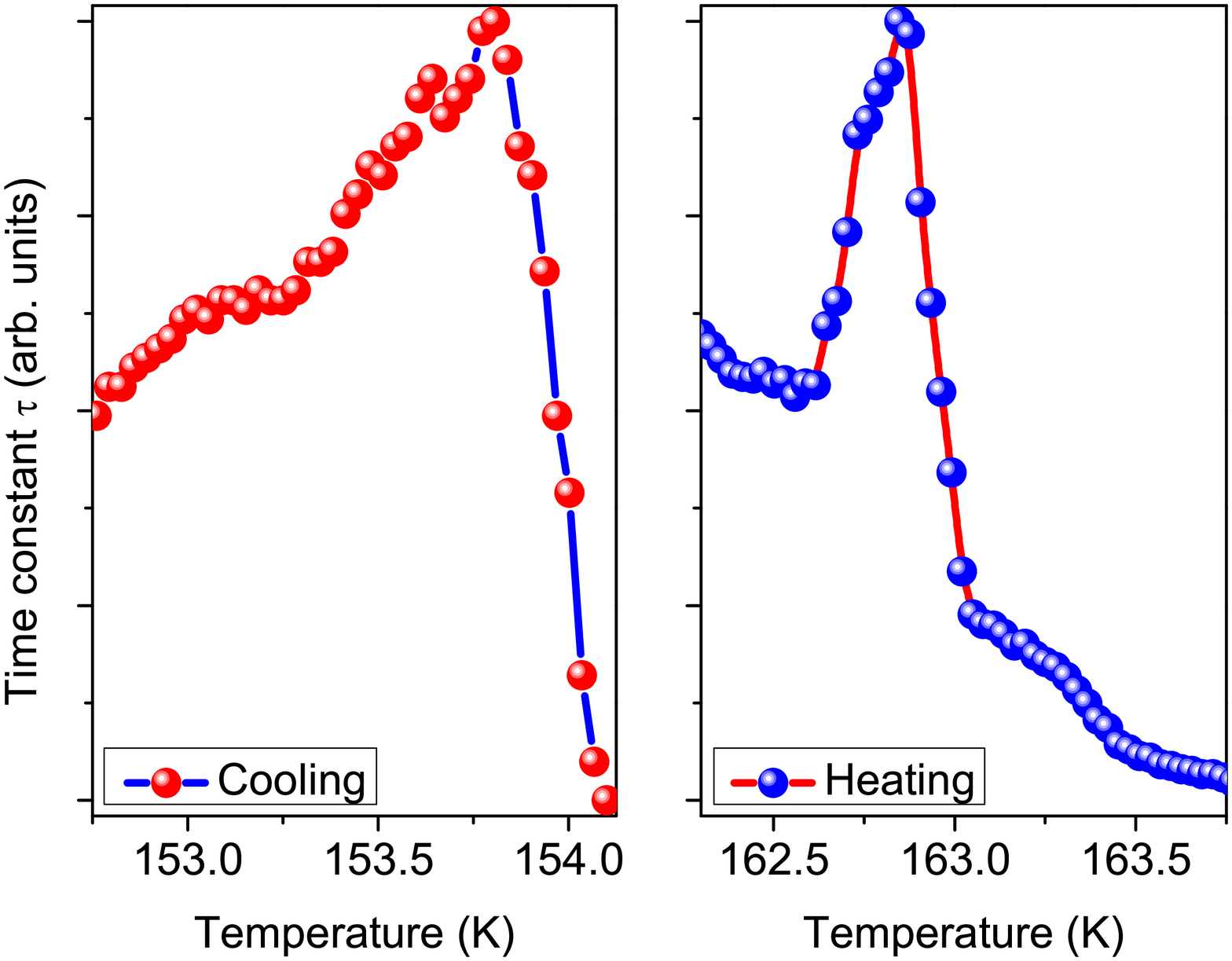}
	\caption{Approximate time constants close to the transition temperature extracted from DTA data during cooling and heating.}
	\label{time_const}
\end{figure}

The same idea is used to determine the phase ordering during the cooling spinodals using cooling FORCs.

For a particular reversal temperature T$_w$, the time-dependent insulator fraction can be extracted from Fig. \ref{FORC} (c) (dashed line in Fig. \ref{FORC} (c). The phase ordering curves, similar to Fig. 8 (supplement), can be constructed from inferring how the insulator fraction varies with the wait time t$_w$ for a given temperature and are plotted in Fig. 10 (supplement). However we found that the inferred time constants are almost an order of magnitude larger than those in Fig. 4 (Main text). There are two reasons for this. Firstly the quench rates are themselves are finite and these measurements were much more difficult than those carried out to generate Fig. 8 (supplement) because they not only required reaching a target temperature and rapidly settling down at that temperature, but also a quench back to T$_s$ after a wait time t$_w$. As can be seen in Fig. 8 (supplement), there is a rapid phase-ordering in the initial $5-10$ seconds. It is especially this initial data which is very imprecise or lost in these thermal analysis measurements as the temperature stabilization time is of the same order. A larger phase ordering time is thus inferred. Secondly, the calculation of the fraction of the latent heat as a function of (T$w$, t$_w$) requires background subtraction, which can be slightly ambiguous when the latent heat is small. Moreover, the goodness of fit in Fig. \ref{FORC} (c) (supplement) is directly related to the goodness of the decay curve.

For these reasons, these latent heat measurements must be considered very rough. Nevertheless, the peaks around the spinodals are in striking qualitative agreement with our other measurements.

\section{Ruling out the effect of temperature fluctuations on the resistance noise} While the temperature stability [Fig 2 (supplement)] during the resistance time series measurements at the two spinodals was essentially limited by the resolution of the temperature controller, it is still worth ruling out that the observed noise is not simply due to the temperature fluctuations.
\subsection{Cross-correlation}
The most direct way to probe this question is to simply compute the cross-correlation between the two time series and quantify the extent to which the measured resistance noise is affected by the temperature fluctuations. The Pearson correlation coefficient $\rho_{_{T,R}}$ provides one such measure. $\rho_{_{T,R}}$ is simply the covariance of the two variables normalized by their standard deviations and in our context may be defined as

\renewcommand\theequation{S3}
\begin{equation}
\rho_{_{T,R}} = {\int_{t_i}^{t_f} dt\, \delta S(t) \cdot \delta T (t) \over \sqrt{\int_{t_i}^{t_f} dt\, \delta T^2 (t)}\sqrt{\int_{t_i}^{t_f} dt\, \delta S^2(t)}}
\end{equation}
Here $\delta S(t)$ is the measured signal and $\delta T(t)$ is the time series of the temperature fluctuations.
$|\rho_{T,R}| \in [0,1]$  with $1$ indicating perfect (linear) correlation and $0$ indicating no correlation. We can equivalently define this correlation in the frequency domain---by the Parseval's theorem the power contained in each of these three integrals is the same in the time and frequency domains, i.e.,

\renewcommand\theequation{S4}
\begin{equation}
\rho_{_{T,R}} = {{1\over 2}\int_{\nu_l}^{\nu_u} d\nu \left[\delta \tilde{T}^*(\nu)\cdot \delta \tilde{S}(\nu)+\delta\tilde{T}(\nu)\cdot \delta \tilde{S}^*(\nu)\right] \over  \sqrt{\int_{\nu_l}^{\nu_u} d\nu \left|\delta\tilde{T}(\nu))\right|^2}\sqrt{\int_{\nu_l}^{\nu_u} d\nu \left|\delta \tilde{S}(\nu))\right|^2}}
\end{equation}
$\delta\tilde{ S}(\nu)$ and $\delta\tilde{T}(\nu)$ denote the Fourier transforms at frequency $\nu$. Taking $\nu_l=0.003$ Hz and $\nu_u=0.3$ Hz, which is the frequency window for which the PSD slopes were calculated in Fig 2 (main text) and what approximately corresponds to the bandwidth of the detrended time series used in the rest of the paper, we find that $\rho_{_{T,R}} \simeq 0.035$ for the measurement at 162 K (heating) in a time series with 120,000 points and $\rho_{_{T,R}} \simeq 0.001$ for the measurement at 153 K (cooling) in the time series with 100,000 points. Thus the correlation in both cases is below 5\% and safely ignored.

What this roughly means is that if the measured signal $\delta S(t)$ is affected by the normalized temperature fluctuations $\delta T(t)$ in the following fashion

\renewcommand\theequation{S5}
\begin{equation}
\delta S(t)\simeq \alpha \cdot \delta S_T(t) + (1-\alpha) \cdot\delta S_R(t), 
\end{equation}
where 
$S_R(t)$ is the true signal and $S_T(t)$ is the spurious component arising out temperature fluctuations $\delta T(t)$, then the fraction $\alpha$ of the measured signal arising from the temperature fluctuations is below 5\%.

\subsection{Temperature coefficient of resistance under different conditions}
While the correlation calculation settles the issue, it is still interesting to understand why this correlation is so small. Physically, we would require

\renewcommand\theequation{S6}
\begin{equation}
{1\over R}\left| {dR\over dT}\right|\delta T << {|\delta R|\over R}
\end{equation}
Here $R$ is the average sample resistance at the temperature of interest, ${dR\over dT}$ temperature coefficient of resistance at the given temperature, $\delta T$ the magnitude of the temperature fluctuations and $|\delta R|$ the magnitude of the noise.

While the change in the sample resistance with temperature at the transition points is very large under a monotonic temperature ramp ($|{1\over R} {dR\over dT}| \sim 2$ K$^{-1}$), luckily the slope is dramatically reduced when one makes small amplitude cyclic excursions around a given point in the metastable phase.  The temperature coefficient of the change in resistance ($|{1\over R} {dR\over dT}|$) is related to the magnitude of the excursion around a given set point as well as the thermal history.

\renewcommand\thefigure{11 (supplement)}
\begin{figure}[h!]
	\begin{center}
		\includegraphics[scale=0.3]{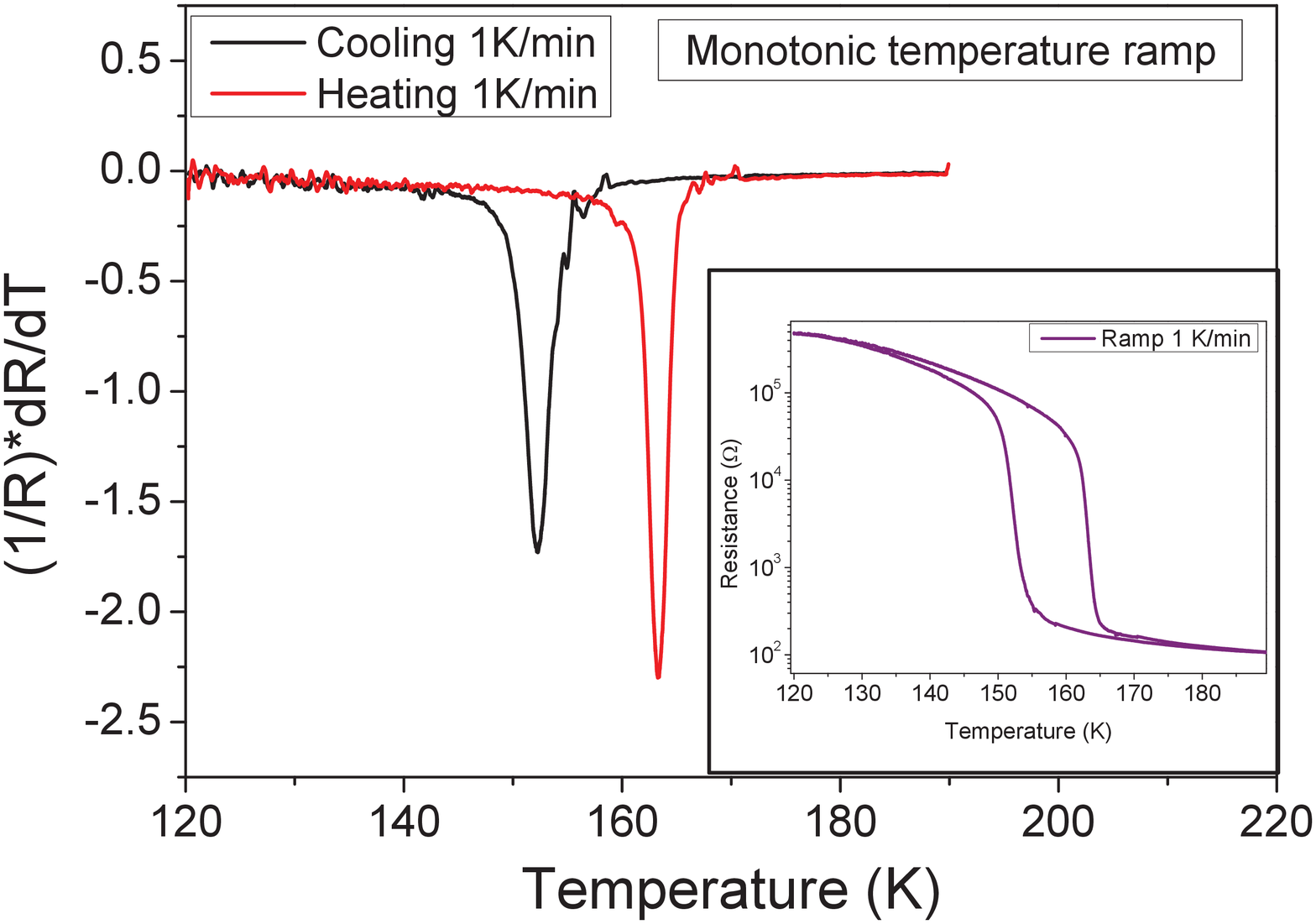}
		\caption{The normalized temperature coefficient of resistance $|{1\over R} {dR\over dT}|\approx 2$ K$^{-1}$ under monotonic temperature ramp. As the next figure shows, this value is greatly reduced when the temperature has a small amplitude fluctuation around a set point. }
		\label{Measurement_setup}
	\end{center}
\end{figure}

\renewcommand\thefigure{12 (supplement)}
\begin{figure}[h!]
	\begin{center}
		\includegraphics[scale=0.3]{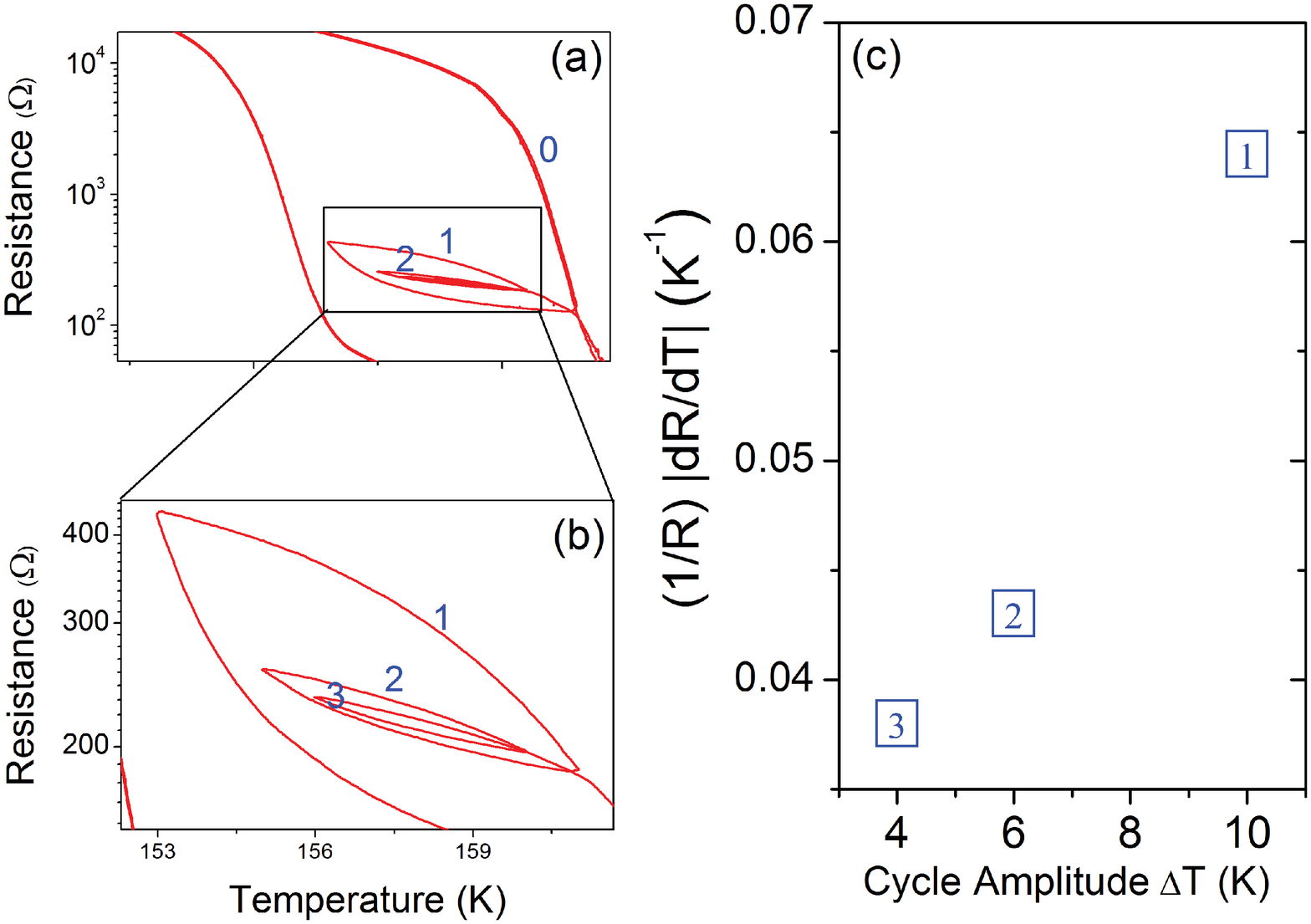}
		\caption{(a) Sample resistance under cyclic temperature variation with the cycle of different amplitudes. (c) $|{1\over R} {dR\over dT}|$ monotonically decreases with the amplitude of temperature oscillation. When $\Delta T$ is $4$ K, the amplitude is reduced to about $0.04$ K$^{-1}$ which is a factor of about 30 lower than the value under a monotonic ramp (marked as `0' in (a)) where $|{1\over R} {dR\over dT}| \approx 1$-$2$ K$^{-1}$  }
		\label{Measurement_setup}
	\end{center}
\end{figure}
Since our noise measurements involve stabilizing the temperature over tens of minutes, it is reasonable to consider the $dR/dT$ under conditions when the temperature goes back and forth around the given set point, rather than what it is under the condition of a monotonic ramp. While we had initially started our measurements by making slow temperature ramps, it was soon realized that this method was untenable for V$_2$O$_3$ due to the large $dR/dT$ and the very sharp transition.

Figure 12 (supplement) shows variation of the normalized slope $|{1\over R} {dR\over dT}|$ in the return-point-memory experiment of Fig. 1 (b) (main text). It can be seen that the slope is a strong function of the amplitude of the excursion in the metastable phase. When the temperature fluctuation is about $4$ K, the slope is only $0.04$ k$^{-1}$. Note that the amplitude of fluctuations in our experiments is nearly three orders of magnitude smaller at $\sim 5$ mK [Fig. 2 (supplement)] and thus even if we take $|{1\over R} {dR\over dT}|\approx 0.01$ K$^{-1}$, a temperature fluctuation of $\delta T \sim 5\times 10^{-3}$ K yields
\begin{eqnarray}
{1\over R}\left| {dR\over dT}\right|\delta T \sim 5\times 10 ^{-5}\nonumber
\end{eqnarray}
Note that ${|\delta R|\over R}$ is already depicted in Fig. 2 (main text) and  ${|\delta R|\over R} \sim 1\times  10^{-3}$.

Thus the condition
\begin{eqnarray}
{1\over R}\left| {dR\over dT}\right|\delta T << {|\delta R|\over R}\nonumber
\end{eqnarray}
is satisfied.

We thus have a qualitative understanding for why the degree of temperature control in our experiments can be considered good enough to not affect the resistance noise.

\section{Spinodals as bifurcation points and the role of fluctuations}
It is perhaps worth pointing out for completeness that, in the language of nonlinear dynamics, phase transitions correspond to bifurcations \cite{Strogatz, Gilmore}. The thermodynamic spinodals that we discuss here also appear in the more general setting of {\em catastrophe theory} \cite{Strogatz, Gilmore, Scheffer-Review}.

The dynamical systems description of bifurcations \cite{Strogatz, verma_dynamo} closely parallels the thermodynamic mean-field theory in terms of the Landau free energy \cite{chaikin-lubensky}. The  continuous or second-order transitions are identified with the supercritical bifurcations \cite{verma_dynamo} and the Landau $\phi^6$ potential used to describe an abrupt phase transition [shown in Fig. 1 (c) (main text)] is a standard manifestation of the pitchfork subcritical bifurcation in the context of nonlinear dynamical systems \cite{Strogatz, Gilmore}. As a consequence, systems undergoing the subcritical bifurcation also display hysteresis and metastability. Subcritical bifurcations have been extensively studied in a variety of (nonthermo-)dynamical systems---e.g., Ref. \cite{Jung_Gray_Roy_Mandel} discusses some dynamical aspects of the lasing instability, Ref. \cite{verma_dynamo} discusses plasma dynamos relevant in astrophysical contexts, and Ref. \cite{Scheffer-Review} discuss such bifurcations or `tipping points' in context of various complex systems.

Furthermore, it not just the potential but the similarity extends also to the dynamical equation. The dissipative Model A (time-dependent Landau-Ginzbug) dynamics \cite{chaikin-lubensky} [Eq. 1 (main text)] is also often used to describe the associated dynamics (with and without noise) \cite{Strogatz, Gilmore} for bifurcations in non-thermodynamic systems. It thus follows that such gradient dynamical systems (where the restoring force is proportional to the gradient of the generalized potential/free energy and there is no inertial term corresponding to the second time derivative) should also exhibit critical slowing down \cite{Jung_Gray_Roy_Mandel} in phase-ordering as the system's resilience (which is the inverse of the order parameter susceptibility) \cite{Scheffer-Review, Gilmore} vanishes around the spinodals. 

Going beyond deterministic dynamics, we have argued in the text that it seems reasonable to expect that the {\em fluctuations} in the relevant coordinate (order parameter) should also exhibit slowing-down and anomalously large variance as one approaches the spinodal bifurcation. This fact has been also been much celebrated in the catastrophe theory literature \cite{Gilmore, Scheffer-Review}. While the arguments presented are exactly along the lines of the theory critical point fluctuations, the question of fluctuations in the context of the spinodals is a delicate matter that had not so far been experimentally resolved. 

Spinodals can only be defined in the limit where the mean field theory is exact and \cite{binder_rpp} and the mean field description of course is arrived at by ignoring fluctuations. The relationship between the spinodal points and fluctuations is thus antithetical and has been doubtful if the manifestations of the spinodal singularity could ever seen in a real thermodynamic system \cite{binder_rpp}.

Theoretical work over especially over the last two decades has carefully addressed this issue in context of various thermodynamic models using both field-theoretic and computational techniques \cite{Gunton-Yalabik, Unger-Klein,Gulbahce, Klein_fluctuations, Gorman-Rikvold-Novotny, Mori-Miyashita-Rikvold}. It has been anticipated that even for systems with (long but) finite-ranged interaction, even though the spinodal should be technically inaccessible, fluctuations may be suppressed enough such that the one characteristic aspect of the mean-field physics of abrupt phase transitions, namely hysteresis, may always be manifest. The effect of the finite range of the interaction is that the divergence is broadened and the susceptibility remains finite at the transition. Correspondingly the lifetime of the metastable phase close to the spinodal is also finite \cite{Gorman-Rikvold-Novotny}. Interestingly, it has been shown that the absence of a true divergence can be thought of as the magnetic field (or temperature) at the point of transition picking up an imaginary component, with the finiteness of the range of the interactions being akin to a finite size effect \cite{Stephanov, Gulbahce}. Such finite-size effects for models with long-range interactions have also been studied in the context of the relaxation dynamics \cite{Mori-Miyashita-Rikvold}. In summary, despite the critical features only approximately manifesting at the transitions, such transitions are expected to be qualitatively very different from the usual abrupt (`first-order') transitions and closer to the critical transitions because one should still observe a rapid growth in fluctuations and the relaxation time scales \cite{Klein_fluctuations}.

Our experimental observations do indeed seem to support this general idea of such mean-field `universality class' \cite{Gulbahce, Miyashita_PRB2009} (or a zeroth order transition \cite{Gilmore}). Although previously spin crossover \cite{Miyashita_PRB2009} and martensite materials \cite{Klein_Nucleation_elastic} have been identified as candidates for this mean-field universality class, this is the first experimental study where the critical slowing down and critical opalescence has been sought and so clearly observed in an abrupt phase transition.


\begin{references}
\bibitem{FN:Spinodal_lightScattering}We must emphasize the difference from the so-called `pseudospinodal' hypothesis \cite{Debenedetti}, where the spinodal line is an infinitesimal extrapolation in the regions of the phase diagram around the critical end-point. Contrastingly our interest is in systems where the APT is strong, there is no critical point in the vicinity, and the spinodals are accessed only after deep supersaturation such that the transition is of zeroth order.
\bibitem{FN:annealed-disorder}However note that avalanche-mediated dynamics seems also derivable within the Landau-Ginzburg scenario \cite{Perez-Reche, Gil-Sornette} by introducing two interacting fields evolving on different time scales. Annealed disorder generated in the first few thermal cycles \cite{Perez-Reche} can lead to stick-slip dynamics of avalanches \cite{Perez-Reche, Gil-Sornette} and the Model A dynamics described here may amount to an effective description, with one variable integrated out. This picture is consistent with our experience, and what is also well-known otherwise, that both polycrystalline samples as well as epitaxial films require a fair amount of training before settling down to reproducible behavior.
\end{references}
\end{document}